\documentclass[reprint,aps,prlCCh,twocolumn,superscriptaddress]{revtex4-1}
\usepackage[utf8]{inputenc}

\usepackage{graphicx}
\usepackage{xcolor}
\newcommand{\bk}{{\mathbf k}}
\newcommand{\bs}{{\mathbf \sigma}}
\usepackage{amsmath,mathtools}

\usepackage{ifpdf}
\usepackage{changes}

\begin{document}
\title{Quantum Anomalous Hall Effect in \\Perfectly Compensated 
Collinear Antiferromagnetic Thin Films}

\author{Chao Lei}
\affiliation{Department of Physics, The University of Texas at Austin, Austin, TX 78712}

\author{Olle Heinonen}
\affiliation{Materials Science Division, Argonne National Laboratory, Lemont, IL 60439, USA}

\author{R.~J.~McQueeney}
\affiliation{Ames Laboratory, Ames, IA, 50011, USA}
\affiliation{Department of Physics and Astronomy, Iowa State University, Ames, IA, 50011, USA}

\author{A. H. MacDonald}
\affiliation{Department of Physics, The University of Texas at Austin, Austin, TX 78712}

\begin{abstract}
We show that the quantum anomalous Hall effect almost always occurs in magnetic 
topological insulator thin films whenever the top and bottom
surface layer magnetizations are parallel, independent of the interior layer magnetization configuration.  
Using this criteria we identify structures that have a quantum anomalous Hall effect 
even though they have collinear magnetic structures with no net magnetization, and discuss 
strategies for realizing these interesting magnetic states experimentally.  
\end{abstract}

\date{\today}

\maketitle

\textit{Introduction---}
The anomalous Hall effect (AHE) was observed \cite{Hall1879,Hall1881} 
already in the 19th century, but understood quantitatively only recently \cite{Nagaosa2010}.  The discovery of the 
quantum Hall effect \cite{QHE_1980,QAH_1986}, and its interpretation \cite{TKNN} in terms of momentum space Chern numbers, 
played a role in improving understanding of the AHE by
clarifying why the intrinsic momentum-space Berry curvature contribution \cite{Karplus1954}, which had sometimes 
been controversial, can play an important role. For many classes of materials predictive theories 
of the AHE, including extrinsic skew \cite{smit1955,smit1958} and side-jump \cite{berger1970side} 
effects along with intrinsic contributions \cite{Karplus1954}, are now available.
The theory of the AHE is especially simple in quasi-two-dimensional magnetic insulators,
since it is then purely intrinsic and must be quantized.  

Historically, the AHE has often been assumed to be proportional
to the magnetization, and therefore to be a characteristic of ferromagnets - not antiferromagnets.
Indeed, rigorous symmetry arguments can be used to rule out an AHE in 
antiferromagnets with a combined $\mathcal{TO}$ symmetry, where $\mathcal{T}$ is time reversal
and $\mathcal{O}$ is any unitary symmetry operator - for example a lattice translation operator.  
This argument rules out an anomalous Hall effect
in the common collinear antiferromagnets of bipartite crystals.  However,
AHEs do occur in both noncollinear \cite{Chen2014,Nakatsuji2015,Nayak_2016,Kiyohara_2016,Chen2021,Naka2020,Tsai2020} and collinear \cite{Smejkal2020,Feng2020} antiferromagnets that do not possess a symmetry of this type.
AHEs in antiferromagnets are 
of technological interest because they provide easy access to information 
stored in hysteretic antiferromagnetic order configurations.

The property that the Hall conductivity of any two-dimensional crystal
$\sigma_{xy} = \sigma e^2/h$ is quantized 
was recognized \cite{Haldane1988} as an outgrowth of the topological theory \cite{TKNN} 
of the quantum Hall effect \cite{QHE_1980}.  Non-zero integer values of $\sigma $ can be produced 
not-only by external magnetic fields but also, in the case of the quantized anomalous 
Hall effect (QAHE), by spontaneous time-reversal symmetry breaking.
The QAHE was first realized \cite{Chang2013} experimentally, in work motivated by a theoretical proposal \cite{Yu_2010},
in magnetically doped and ferromagnetically ordered topological insulators.
The possibility of a QAHE in non-collinear, and non-coplanar two-dimensional antiferromagnets
has been thoroughly explored theoretically \cite{Martin2008,Venderbos2012,Ishizuka2013,Chern2014,Ndiaye2019}.
Here we predict that the QAHE also occurs in magnetic topological 
insulator (MTI) thin films with perfectly compensated collinear magnetic order when the top and bottom surface layer magnetizations are parallel, and discuss 
how these magnetic configurations can be realized experimentally.

\ifpdf
\begin{figure}[htp]
\includegraphics[width=0.9\linewidth]{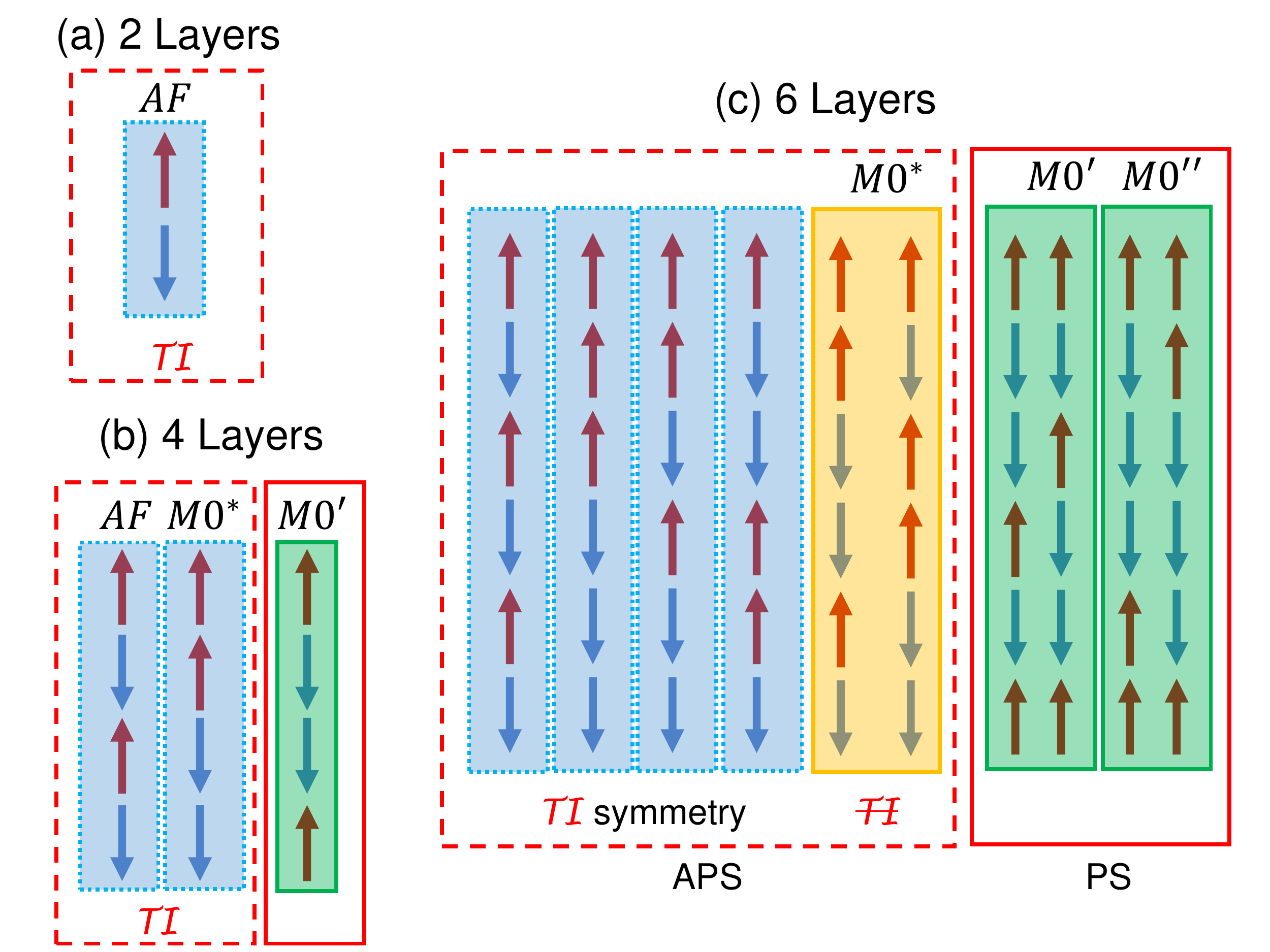}
    \caption{Fully compensated magnetic configurations of 2, 4, and 6 layer MTI thin films.  Only configurations 
    that are distinct under global spin reversal are illustrated.  Odd parity configurations (blue) yield a band 
    Hamiltonian with $\mathcal{TI}$ symmetry and necessarily have Hall quantum integer $\sigma=0$. 
    We find that configurations that are not odd parity, but still have anti-parallel surface layers (APS)
    (orange) almost always have $\sigma=0$, while those with parallel surface layers (PS - green)  
     almost always have $\sigma \ne 0$.}
\label{fig:mag_configuration}
\end{figure}
\fi

\textit{Qualitative QAHE Criteria}---
Mn(Sb$_x$Bi$_{1-x})$X$_4$ thin films consist of van der Waals coupled septuple layers with ferromagnetically
ordered Mn local moments at their centers, and perpendicular-to-plane easy axes \cite{Eremeev2017,Yan2019,Li2020,Zeugner2019,Otrokov2019}.
The ferromagnetic layers are coupled via weak antiferromagnetic 
exchange interactions that act across the van der Waals gap. 
This interesting family of materials has recently attracted both theoretical and experimental interest  \cite{Lei2020,Otrokov_2017,Eremeev2017,Otrokov2019,Zhang2019,Li2019_theory,Chowdhury_2019,Lee2013,Rienks2019,Zeugner2019,Yan2019,Lee2019,Li2020,Otrokov2019_film,Liu2020,Chen2019_Pressure,Deng2020,Deng_2020,Gong2019,Zhang2019_AHC,Li2019,Hao2019,Chen2019,Ge2020,Hu2020,Ding2020,Lee2019,Swatek2020,Eremeev2018,Wu2019,Vidal2019,Klimovskikh2019,Sun2019,Gu_2020,Wimmer2020,Belopolski2017}. 
Compensated antiferromagnetic states with equal numbers of $\uparrow$ and 
$\downarrow$ layers are possible for even layer number $N$. The number of compensated
magnetic configurations is $C(N,N/2) = 2,6,20 \ldots$ for $N=2,4,6 \ldots$. 
Each of these combinations has a time-reversed
partner whose Chern number differs by a sign.
Choosing one member from each time-reversed pair leaves the $C(N,N/2)/2$ configurations, illustrated for 
$N=2,4,6$ in Fig.~\ref{fig:mag_configuration}, to be studied.

Our analysis of MTI thin films is based on a simplified 
couple Dirac cone model \cite{Lei2020} applicable to the Mn(Sb$_x$Bi$_{1-x}$)X$_4$ X = (Se,Te)
family of intrinsic magnetic topological insulators (IMTIs),
QAHEs are expected to be common in odd $N$
uncompensated films, and have been observed
in MBT (MnBi$_2$Te$_4$) thin film with $N=5$ \cite{Deng2020}.  QAHEs have also 
been observed at other film thicknesses \cite{Deng2020,Ge2020,Liu2020} 
when the magnetic configurations is altered 
by applying magnetic fields larger than $\approx 5$ T.
In Fig. \ref{fig:mag_configuration} we classify the magnetic configurations of 
even $N$ compensated moment MTI thin films as either anti-parallel surface layer (APS) or parallel surface layer (PS),  
depending on whether the magnetizations of the 
top and bottom surface layers are anti-parallel or parallel.  
For large even $N$ the number of PS configurations is almost equal to the number of APS configurations \cite{SI}.
Many APS films have odd-parity magnetization configurations in the sense that their magnetizations are reversed when the layer order is reversed.  
The mean-field Hamiltonians of this subset of 
APS films (blue in Fig. \ref{fig:mag_configuration}) can be shown \cite{Lei2021_metamagnetism} to have 
$\mathcal{TI}$ symmetry and hence $\sigma=0$.  We find numerically that even APS films that do not have this symmetry (orange in Fig. \ref{fig:mag_configuration}) almost always have $\sigma=0$.  On the other 
hand PS magnetic configurations (green in Fig. \ref{fig:mag_configuration}) 
often have $\sigma \ne 0$, even though their moments are perfectly compensated.  

We start by examining the $N=2$, $N=4$ and $N=6$ cases in detail.  $C(N,N/2)/2=1$ for $N=2$, leaving one configuration
to be studied.  Since this configuration has odd parity, the mean-field Hamiltonian has 
$\mathcal{TI}$ symmetry and we know without calculation that Berry curvature vanishes and $\sigma=0$.
The thinnest PS configuration occurs at $N=4$.  The 3 configurations of $N=4$ thin film in 
Fig.~\ref{fig:mag_configuration}(b) are labelled $AF$, $M0^{\ast}$ and $M0'$.  Among these only the 
$MO'$ ($\uparrow \downarrow \downarrow \uparrow$) state has PS configuration and thus can host a 
QAHE state. For $N=6$, four of the ten illustrated configurations have
$\mathcal{TI}$ symmetry and thus zero Berry curvature and $\sigma=0$.
To determine whether or not the two remaining APS magnetic configurations 
(denoted as $M0^{\ast}$) and the four PS configurations (denoted as $M0'$ and $M0"$) in Fig. \ref{fig:mag_configuration} (c)) support QAHE states, it is necessary to examine the electronic structure more closely.

\ifpdf
\begin{figure}[htp]
\includegraphics[width=0.95\linewidth]{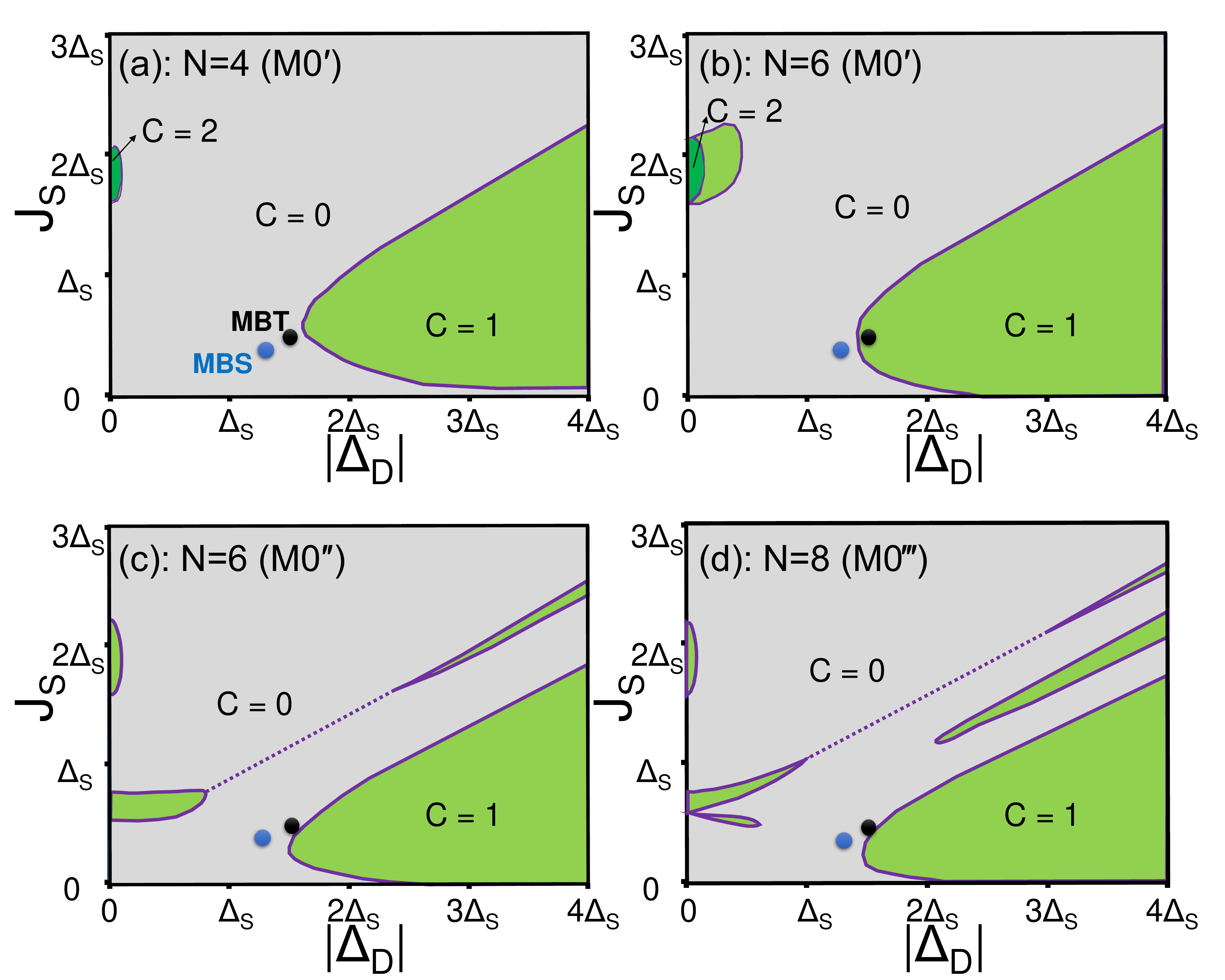}
    \caption{MTI thin film phase diagrams for a variety of fully compensated PS magnetic configurations.
    The $x$ and $y$ axes are the ratio of the interlayer to intralayer hybridization $|\Delta_D|/\Delta_S$ and 
    the ratio of exchange and hybridization parameters $J_S/\Delta_S$. 
    The phase boundary is also sensitive to $\delta \equiv J_D/J_S$, the ratio of the exchange coupling to adjacent layer moments to the exchange coupling to same layer moments, which is fixed in these plots at $\delta = 0.8$, 
    the value estimated for MBT.  Phase diagrams for larger and smaller values of 
    $\delta$ are included in the supplementary material \cite{SI}.
    (a) $N=4$ M0$'$ state;
    (b) $N=6$ M0$'$ and (c) $N=6$ M0$''$ state;
    (d) $N=8$ M0$'''$ state with magnetic configuration ($\uparrow \downarrow \downarrow \downarrow \downarrow \uparrow \uparrow \uparrow$);
The light green regions of the phase diagram have Chern number magnitude $|C| = 1$, 
whereas the dark green regions have $|C| = 2$ and the grey regions $C=0$. Model parameters estimated for MnBi$_2$Se$_4$ and MnBi$_2$Te$_4$ 
at temperature $T=0$ are marked by blue and black dots respectively.
}
\label{fig:phase_diagram}
\end{figure}
\fi

\textit{Model Calculations}--- 
We employ a low-energy phenomenological band model, discussed in detail in \cite{Lei2020},
with Dirac cones on both surfaces of each septuple layer: 
\begin{equation}
\begin{split}
   H = & \sum_{\bk_{\perp},ij} \Big[\Big( \, 
   (-)^i  \hbar v_{D}  (\hat{z} \times \bs) \cdot \bk_{\perp} + m_{i} \sigma_z \Big) \delta_{ij}   \\
   & + \Delta_{ij}(1-\delta_{ij} ) \Big] c_{\bk_{\perp} i}^{\dagger} c_{\bk_{\perp} j}.
\end{split}
\end{equation}
Here $i$ and $j$ are Dirac cone labels with even integers reserved for septuple layer bottoms 
and odd for layer tops, $ \hbar$ is the reduced Planck constant, $v_{_D}$ is the velocity of the Dirac cones,
and $\Delta_{ij}$ is the hopping amplitude between the i$^{th}$ j$^{th}$ Dirac cones.
Only the four largest model parameters, estimated by fitting to DFT calculations, 
are retained in our calculations: hopping between the surface Dirac cones in the same layer($\Delta_S$),
nearest neighbour hopping between adjacent layers ($\Delta_D$), and two exchange coupling parameters.
The exchange coupling parameter $m_i \equiv \sum_{\alpha} J_{i\alpha} M_{\alpha}$ 
where $ \alpha $ is a layer label and $M_{\alpha} = \pm 1$ specifies the sense of magnetization on layer $\alpha$.  
We retain exchange coupling $J_S$ to the magnetization in the 
same septuple layer and near-neighbor exchange coupling $J_D$ to the magnetism in the adjacent septuple layer.  

Fig. \ref{fig:phase_diagram} contains two-dimensional 
$|\Delta_D|/\Delta_S$-$J_S/\Delta_S$ topological phase diagrams 
calculated with $\delta \equiv J_D/J_S=0.8$ - its MBT value \cite{Lei2020}.
This figure includes phase diagrams for several fully compensated PS magnetic configurations:
4-layer $M0'$, 6-layer $M0'$ and $M0''$, and 8-layer $M0'''$. 
In the phase diagrams light green regions represent 
quantum anomalous Hall states with Chern number $C= 1$, 
the dark green regions represent $C= 2$, and the gray regions represent normal insulators.
Quantum anomalous Hall (QAH) states are common in the bottom right regions of these
phase diagrams, where $\Delta_D$ is large enough to yield TI states in the absence of 
magnetism and $J_{S}$ is small enough that the exchange fields perturb the non-magnetic 
TI state weakly.  The model parameters estimated for MBT are close to the phase boundaries 
between QAH and trivial states because the exchange interactions are relatively weak and 
because these materials are barely topological in the sense \cite{Lei2020} that $|\Delta_D/\Delta_S|$ is not much larger than one.  

It is instructive to examine the $J_S=0$ and $\Delta_D=0$ lines in the phase diagrams more closely.
We do this in Fig. \ref{fig:gaps} by plotting thin film energy gaps {\it vs.} $\Delta_D$ at $J_S=0$ and 
{\it vs.} $J_S$ at $\Delta_D$=0.  In Fig. \ref{fig:gaps} (a) we see that large values of $\Delta_D$ increasingly 
isolate the top and bottom surface Dirac cones and decrease the amplitude for tunneling between them across the bulk
of the film. The surface isolation property at large $\Delta_D$ can be understood qualitatively  
by examining the bilayer limit of the Dirac cone model, for which the band gap at $J_S=0$ 
is $E_g = \sqrt{\Delta_D^2 + 4\Delta_S^2} - |\Delta_D|$, which goes to 0 whenever $\Delta_D \rightarrow \infty$.
This property explains the proximity of the QAH region to the $J_S=0$ line at 
large $|\Delta_D|$, since very weak exchange is then sufficient to induce a level crossing between 
the surface states.

Along the $\Delta_D=0$ line, whose gaps are plotted Fig. \ref{fig:gaps} (b),
each septuple layer is an isolated two-Dirac-cone two-dimensional electron system that contributes a quantum 
unit to the anomalous Hall effect when its exchange coupling strength exceeds $\Delta_S$.
The isolated septuple layer Hamiltonian is 
\begin{equation}
    H_{SL} = \begin{pmatrix}
m_t & v_D k_- & \Delta_S & 0 \\
v_D k+ & -m_t & 0 & \Delta_S \\
\Delta_S & 0 & m_b & -v_D k_- \\
0 & \Delta_S &  -v_D k+ & -m_b\\
\end{pmatrix},
\end{equation}
where $k_{\pm} \equiv k_y \pm i k_x$, and $m_{t/b}$ are top and bottom surface Dirac masses.
On the outside surfaces $m_{t/b}$ equals $\pm J_S$, whereas on the interior surfaces $m_{t/b}$ can equal 
$\pm (J_S+J_D)$ or $\pm(J_S-J_D)$ depending on the magnetic configuration.
The eigenvalues of this Hamiltonian are 
\begin{equation}\label{eq:isolated}
    E = \pm \frac{1}{2} \sqrt{4v_D^2 |k|^2 + \Big( m_+ \pm \sqrt{m_-^2 + 4\Delta_S^2} \Big)^2},
\end{equation}
where $|k| = \sqrt{k_x^2 + k_y^2}$ and $m_{\pm} \equiv m_t \pm m_b$.
From Eq. \ref{eq:isolated} we see that the gaps are determined by the band energies at the two dimensional 
$\Gamma$ point.  Using Eq.~\ref{eq:isolated} it is easy to determine the Hall conductance 
contributed by each septuple layer along the $\Delta_D=0$ line in any magnetic configuration.
The $C=2$ regions along the $\Delta_D=0$ line in Fig. \ref{fig:phase_diagram}
appear when the surface septuple layers have entered QAH states, but the interior septuple layers still have zero Chern number.

\ifpdf
\begin{figure}[htp]
\includegraphics[width=0.95\linewidth]{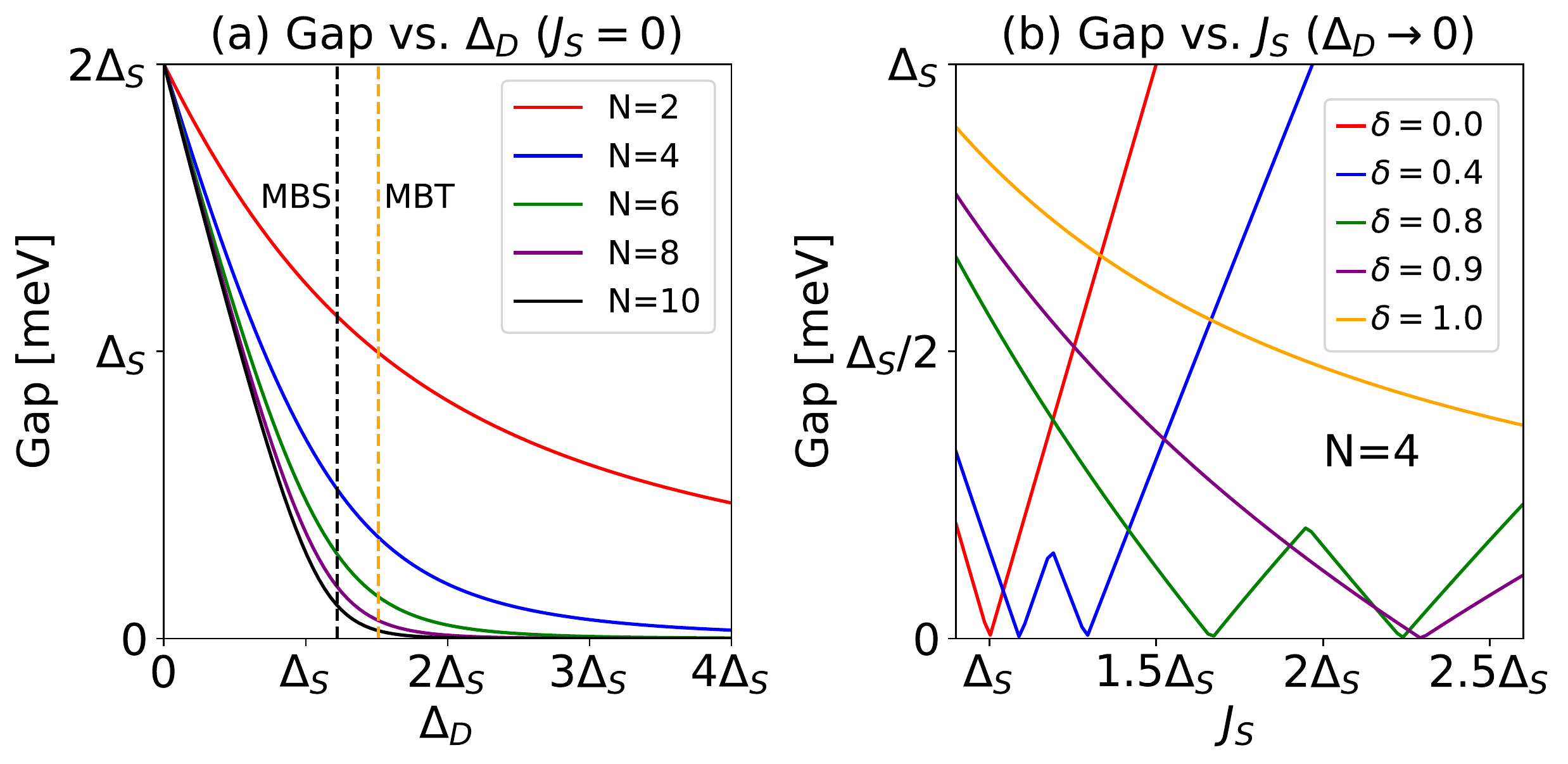}
    \caption{Gaps {\it vs.} hybridization between different septuple layers $\Delta_D$ and exchange splitting $J_S$.
    The quantized Hall conductance can change value only when gaps close.
    (a) dependence on $\Delta_D$ at $J_S=0$ for several thin films thicknesses.
    (b) dependence on $J_S$ at $\Delta_D=0$ for the four-layer thin film with the parallel surface magnetic configuration.  
    Several different values of $\delta \equiv J_D/J_S$ are considered. 
    For $\delta = 0$ and $\delta = 1$ no topological phase transition occurs as a function of $J_S$, in the former case because each septuple layer contributes the same sign of Hall conductivity as its spin magnetization, and in the latter case because
    topological transitions are absent.  For other values of $\delta$ topological phase transitions occur between
    $C = 0$ and $C = 2$ states.
    }
\label{fig:gaps}
\end{figure}
\fi

The sensitivity of the phase diagram to $\delta$ is greatest at small $\Delta_{D}$.
When $\delta$ and $\Delta_D$ both vanish each septuple 
layer is driven into a QAH state when $J_S > \Delta_S$ with the Chern number sign 
depending on the direction of magnetic moment in that layer.
It follows that for all perfectly compensated configurations the total Chern number vanishes in this limit. 
When $\delta \to 1$ on the other hand a variety of different cases must be distinguished.
Consider, for example, the top septuple layer when it is isolated by setting $\Delta_{D} \to 0$.
For a $\uparrow \uparrow \cdots$ configurations
the energies at $\Gamma$ are $E = (\pm J_S \pm \sqrt{J_S^2 + 4\Delta_S^2})/2$, whereas for a  
$\uparrow \downarrow \cdots$ configurations $E = (\pm 3J_S \pm \sqrt{J_S^2 + 4\Delta_S^2})/2$.
Similarly for an interior layers with an $\cdots \uparrow \uparrow \downarrow \cdots$ configuration 
$E = \pm J_S \pm \sqrt{J_S^2 + \Delta_S^2}$, whereas for $\cdots \uparrow \uparrow \uparrow \cdots$
configurations $E = \pm 2 J_S \pm \Delta_S$. When level crossings occur as a function of $J_S$, the isolated septuple layer's contribution to the Hall conductivity changes from $0$ to $1$.    The appearance or absence of QAH phases is easily determined by adding the contributions of all layers.  These types of considerations explain the phase transition points along $\Delta_{D}=0$ lines in Fig.~\ref{fig:phase_diagram}, Fig.~\ref{fig:gaps}(b),
and in Fig.~S5 of the supplemental material\cite{SI} which presents phase diagrams for $\delta = 0$ and $\delta=1$.


\textit{Discussion}---
We have shown that magnetic configurations of
Mn(Bi$_x$Sb$_{1-x}$)$_2$X$_4$ multilayer MTI thin films with parallel magnetizations on surface septuple layers 
can have QAHEs even though they have perfectly compensated collinear spin moments.
The thinnest example is a four layer structure with 
interior and exterior layer magnetizations having opposite orientations ($\uparrow \downarrow \downarrow \uparrow$), 
but many more configurations in this category appear in thicker films \cite{SI}. 
The appearance or absence of a QAHE is dependent on the details of electronic structure and 
magnetic interactions, and that dependence is described 
here in terms of the parameters of a simplified Dirac-cone model of the 
electronic structure with hybridization and exchange parameters 
$\Delta_S$, $\Delta_D$, $J_S$ and $J_D$ that predicts 
the phase diagrams in Fig. \ref{fig:phase_diagram}.
It is possible to some extent to move through this phase diagram experimentally by 
varying the choice of chalcogen $X$ or the pnictide fraction $x$ in 
Mn(Bi$_x$Sb$_{1-x}$)$_2$X$_4$, by apply vertical strains, or by increasing temperature to reduce exchange interaction strengths. 
For the case of MBT with $N \ge 6$ and the chosen parameters, all configurations with parallel surface (PS) layer magnetization 
have a QAH phase over a finite interval of temperature
when thermal fluctuations in local moment orientations are assumed to decrease
exchange interaction strengths ($m_{i} \to \xi m_i$ with $\xi \in (0,1)$) as shown in Fig. S4 in the supplemental material \cite{SI}. For magnetic configurations with anti-parallel surface magnetizations, no gap closings
occur as a function of $\xi$, indicating that all remain in their $\xi=0$ topologically trivial states (with zero Chern number) at any temperature.

\ifpdf
\begin{figure}
\includegraphics[width=0.9\linewidth]{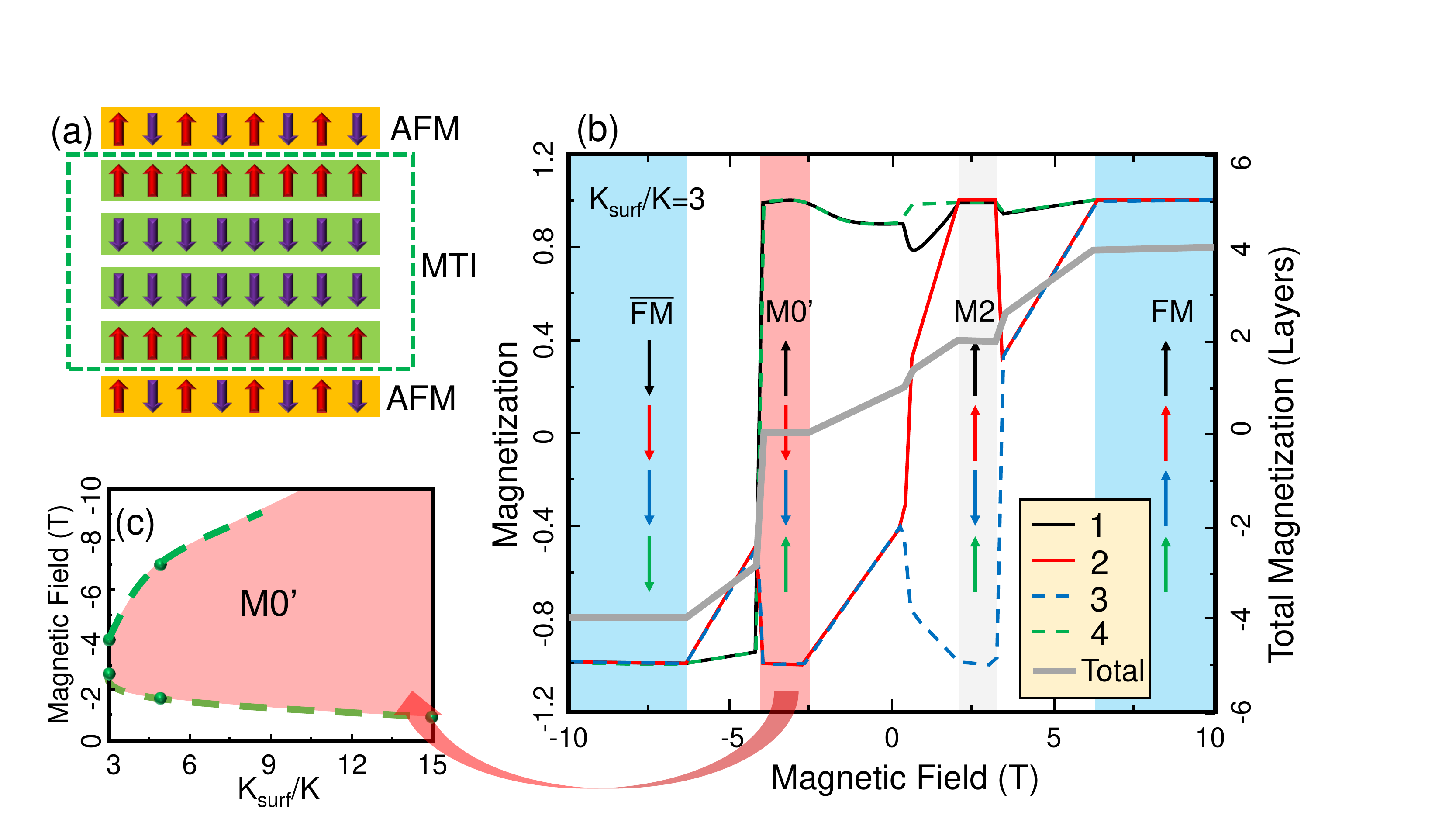}
    \caption{The $N=4$ parallel-surface-layer (PS) fully compensated collinear magnetic configurations can be realized by 
    reversing the orientations of the top or bottom layers relative to the interior layers using exchange bias. 
    The red and purple arrows in (a) represent moment orientations. 
    (b) shows the $z-direction$ total magnetization that is predicted by classical Monte Carlo simulation 
    of a $N=4$ film at the temperature of 0.1 K. The effective surface layer magnetic anisotropy is 3 time the interior layer magnetic anisotropy.
    The shaded regions have collinear magnetic configurations, and the integer portion of the configuration label is the net number of aligned layers.  The PS configuration is labelled M0$'$.
    (c) shows parameter range of surface magnetic anisotropy and magnetic field over which $N=4$ PS states 
    occur when the magnetization is swept toward negative values starting from $B=10$ T.
    }\label{fig:scheme}
\end{figure}
\fi

The set of magnetic configurations that can be realized by cycling magnetic field is dependent on magnetic anisotropy.
The PS configurations of interest here are accessible when the magnetic anisotropy energy is favorable, 
either naturally or as a consequence of intentional
interface engineering. For example, in a previous publication \cite{Lei2021_metamagnetism} we showed that the $N=6$ PS magnetic configuration M0' state in Fig. \ref{fig:mag_configuration}
(referred to there as M0$^{\ast}$) can be reached from the M2$'$ state when the ratio of the single-ion anisotropy 
coefficient $D$ to the interlayer exchange interactions $J$ is sufficiently large.  Specifically $D/zJ$ must be greater than $0.23$,
much larger than the $D/zJ$ ratio of bulk MnB$_2$X$_4$ which is approximately $0.13$.  (The bulk single-ion anisotropy energy of 
MnB$_2$X$_4$ $K \equiv D S^2$, where $S = 5/2$, is approximately $0.17$ meV and corresponds to a value of $SD = 0.07$ meV.  The interlayer exchange interaction $SJ=0.088$ meV and the interlayer coordination number $z=6$.)
Although it seems likely, therefore, that the bulk anisotropy energy is insufficient to reach PS configurations in
MnB$_2$X$_4$, it will nevertheless be interesting to explore other materials with 
similar structures that could very well have more favorable $D/zJ$ ratios, possibly MnBi$_4$Te$_7$ where the interlayer coupling strength is much weaker.

Another strategy that can be used to realize PS magnetic configurations is to engineer the effective 
anisotropy of the surface layers. 
This could potentially be accomplished by exchange-biasing the surface layers, as illustrated in 
Fig. \ref{fig:scheme} (a). Exchange biasing may not be possible for the bottom surface if the choice of 
substrate is very important for the epitaxial growth of 
MBT thin films. However, since MBT can be exfoliated, it is possible to transfer exfoliated flakes onto some other insulating AFM such as the (111) surface of FeO in which the magnetization in each (111) plane is perpendicular to the (111) plane.  There is more 
flexibility in engineering the top surface anisotropy since we may grow an insulating AFM with decent coupling on the top surface. 
Perpendicular exchange bias would provide an added effective uni-directional anisotropy to the surface layers so that the interior layers will switch in an applied magnetic field while the surface layers do not.
This qualitative idea is quantified in Fig. \ref{fig:scheme} (b), which plots the $\hat{z}$-component of 
total magnetization as a function of magnetic field, as the magnetic field is reversed from a saturating 
$B=10$ Tesla field.  For the illustrated case in which the surface layer anisotropy is three times the interior layer anisotropy (assumed to equal the bulk crystal value) the PS Chern insulator configuration is stable over a finite range of magnetic field. It is interesting that
the magnetic-field cycling strategy can work even though a QAH state with vanishing total spin magnetization.   
As illustrated in Fig. \ref{fig:scheme} (c), the Chern insulator configuration with perfect 
spin moment compensation occurs over a wider and wider range of magnetic field as the surface 
layer magnetic anisotropy gets stronger and stronger.  
In principle it is possible to set the exchange bias by applying a large field near the N\'eel temperature of the insulating AFM and then cool down in field to well below the insulating AFM N\'eel temperature (below the so-called blocking temperature for exchange bias). 
For many exchange bias systems, the blocking temperature is much higher than the Neel temperature of MBT. Therefore, in the relevant temperature range here, the
exchange bias on the top and bottom surfaces will then be basically independent of temperature. In this way we can tune the interlayer 
exchange coupling by tuning the temperature in a regime where the exchange bias remains fixed.
PS QAH states are likely easier to realize in thicker films, where conifgurations with this property are abundant,
but the QAH effect itself is more vulnerable \cite{Lei2021_QAH} to unintended external electric fields.  

In summary we have shown that MBT thin films have a QAHE in spite of having perfect spin-moment compensation, possibly supported in some metastable magnetic configurations. Magnetic states with these unusual properties could potentially be valuable for applications, if they could be realized in well controlled materials with room temperature magnetic ordering temperatures.

\section{Acknowledgements} C.L. and A.H.M. were sponsored by the Army Research Office under Grant Number W911NF-16-1-0472. R.J.M. and O.H. were supported by the Center for Advancement of Topological Semimetals, an Energy Frontier Research Center funded by the U.S. Department of Energy Office of Science, Office of Basic Energy Sciences, through the Ames Laboratory under Contract No. DE-AC02-07CH11358. We gratefully acknowledge the computing resources provided on Blues, a high-performance computing cluster operated by the Laboratory Computing Resource Center at Argonne National Laboratory.

\bibliography{mbt}
\end{document}